\documentclass{article}
\usepackage{spconf,amsmath,graphicx}

\usepackage[T1]{fontenc}

\usepackage{graphicx}
\usepackage{amsmath}
\usepackage{url}            % 
\usepackage{xcolor}         % colors
\usepackage[labelformat=simple]{subcaption}
\usepackage{caption}
\usepackage{enumerate}
\usepackage{algorithm}
\usepackage{array}
\usepackage{multirow}
\usepackage{caption}
\usepackage{float}
\usepackage{amssymb}
\usepackage{color,xcolor}
\usepackage{afterpage}
\usepackage{siunitx}
\usepackage{microtype}

\title{A M\lowercase{odel} S\lowercase{tealing} A\lowercase{ttack} A\lowercase{gainst} M\lowercase{ulti}-E\lowercase{xit} N\lowercase{etworks}}
\name{Pan Li,
Peizhuo Lv\textsuperscript{*},
Kai Chen\thanks{*Corresponding Authors.}\textsuperscript{*$\dagger$}\thanks{$\dagger$The IIE authors are supported in part by NSFC (92270204, U24A20236), CAS Project for Young Scientists in Basic Research (Grant No. YSBR-118).},
Shengzhi Zhang,
Yuling Cai,
Fan Xiang,
}
\address{
    SKLOIS, Institute of Information Engineering, Chinese Academy of Sciences, China\\
    School of Cyber Security, University of Chinese Academy of Sciences, China\\
    Department of Computer Science, Metropolitan College, Boston University, USA
}

\begin{document}
%\ninept
%
\maketitle

\begin{abstract}
Compared to traditional neural networks with a single output channel, a multi-exit network has multiple exits that allow for early outputs from the model's intermediate layers, thus significantly improving computational efficiency while maintaining similar main task accuracy. Existing model stealing attacks can only steal the model's utility while failing to capture its output strategy, i.e., a set of thresholds used to determine from which exit to output. This leads to a significant decrease in computational efficiency for the extracted model, thereby losing the advantage of multi-exit networks. In this paper, we propose the first model stealing attack against multi-exit networks to extract both the model utility and the output strategy. We employ Kernel Density Estimation to analyze the target model's output strategy and use performance loss and strategy loss to guide the training of the extracted model. Furthermore, we design a novel output strategy search algorithm to maximize the consistency between the victim model and the extracted model's output behaviors. In experiments across multiple multi-exit networks and benchmark datasets, our method always achieves accuracy and efficiency closest to the victim models.
\end{abstract}

\section{Introduction}
\label{sec:intro}

% To insert a figure: \input{figs/template}
% Or table: \input{tables/template}

%The rapid development of neural networks has witnessed their extensive applications in our daily lives, such as computer vision, speech recognition, natural language processing, etc. However, with continuous demand for higher performance from these applications, the number of parameters and the complexity of models also keep growing, greatly increasing computational cost
With relentless pursuit of higher AI performance, the scale and complexity of models keep growing, greatly increasing computational cost \cite{quantization, quantization2,quantization3}. This poses a significant challenge for model deployment in scenarios with real-time requirements or limited computing power, e.g., mobile or IoT devices \cite{IoTSenario,LightweightIoT1,LightweightIoT2}. Inspired by the fact that canonical samples can produce sufficiently confident results with fewer computation \cite{SDN}, multi-exit networks have emerged as a promising technique to enhance the computational efficiency of models. 

\begin{figure}[ht]
\centering
\includegraphics[width=0.45\textwidth]{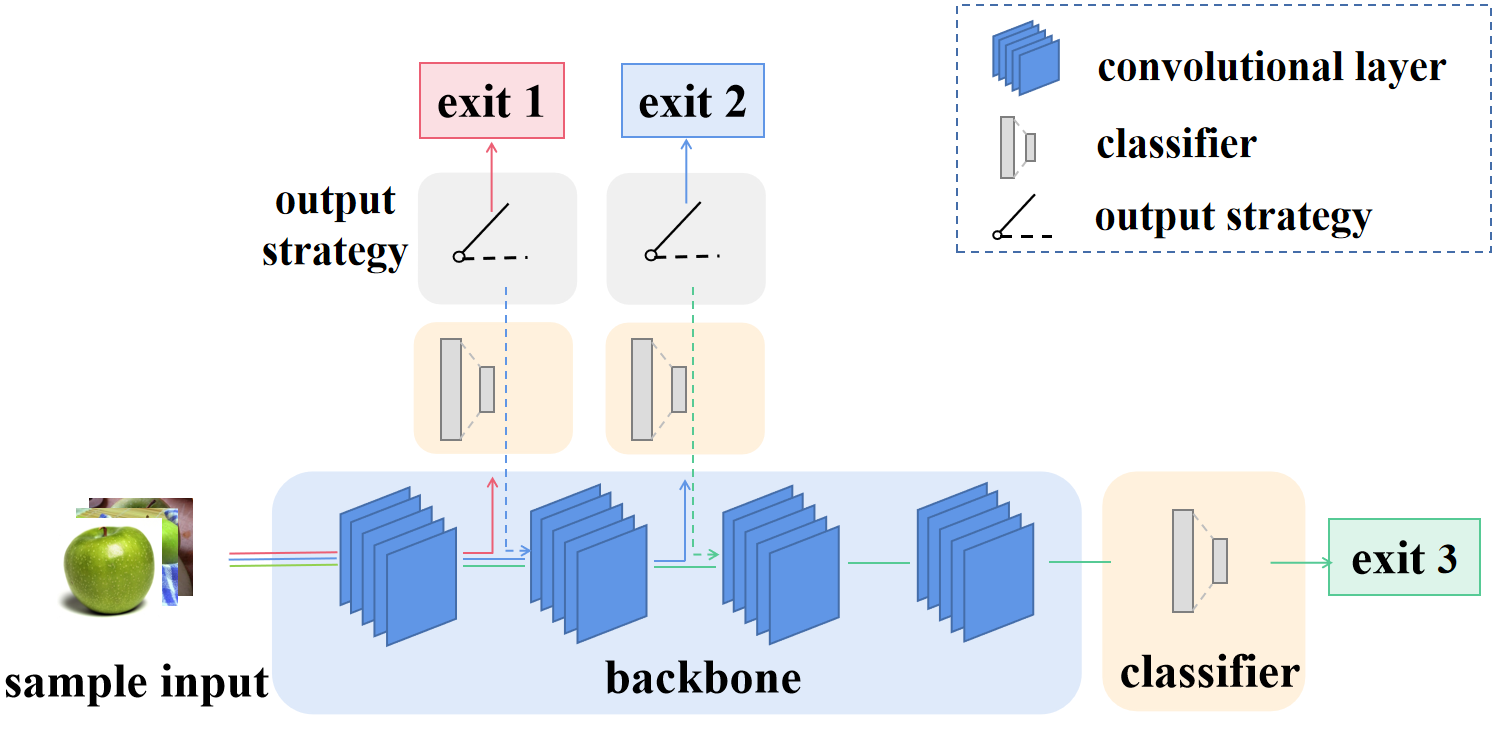}
\caption{Multi-Exit Model Structure.
%\end{tablenotes} 
}
\label{fig:multi-exit-structure}
\vspace{-15pt}
\end{figure}

Unlike traditional neural networks with a fixed execution path, multi-exit networks \cite{multi-exit_architecture1,multi-exit_architecture2} offer multiple classifiers as potential exits, as shown in Fig.\ref{fig:multi-exit-structure}. This architecture allows multi-exit networks to output in advance based on output strategies, thereby avoiding unnecessary computational costs. With a good output strategy, a multi-exit network can allocate fewer computational resources for ``simple" samples and more for ``complex" samples, thereby greatly improving the computational efficiency of the model while maintaining similar recognition accuracy. The output strategy for a multi-exit network is typically a set of predefined thresholds. When the model's confidence in the current exit exceeds the corresponding threshold, it indicates that the prediction is sufficiently confident and can be output early.

Due to the advantages of fast inference \cite{application1,application2, application3} and energy-efficient computation \cite{research1,research2,research3}, stealing multi-exit networks is very appealing to attackers. 
%The characteristics of fast inference and efficiency calculation strongly attract attackers to launch model stealing attacks against such valuable networks
However, we find that the unique structure and output strategies of multi-exit networks render existing model stealing methods ineffective since they only focus on stealing the model's classification functionality but ignore the output strategy. This leads to a significant downgrade in the computational efficiency of the extracted model, thus losing the advantage of multi-exit networks. 
%For instance, our experiments demonstrate that the extracted models trained via KnockOff\cite{KnockoffNets}, Noise\cite{Noise} and ES attack\cite{Es-attack} averagely require 44\%, 48\%, and 47\% more computational resources than the victim models. This is unacceptable in scenarios with limited computational resources and high real-time requirements. 
%Furthermore, only 29\%, 26\%, and 25\% of the samples exhibited similar output behavior as that of the victim model, which indicates existing methods cannot effectively learn the victim model's output behavior.

In this paper, we propose a novel model stealing attack against multi-exit networks that steals not only the model's classification functionality but also the output strategy, making the extracted model approximate the output behavior of the victim model as closely as possible. 
%We employ Bayesian Change-point Detection (BCD) to estimate the output strategy by statistically analyzing the runtime of input samples on the victim model. 
We first employ Kernel Density Estimation (KDE) \cite{KDE1, KDE2} to estimate the time threshold of all the exits of the victim model and predict exit numbers for each query sample. 
Then, we introduce performance loss and strategy loss to train the extracted model, enabling it to learn the corresponding function of the victim model. Finally, we propose a new output strategy search algorithm to determine the optimal output strategy that maximizes the consistency between the extracted model and the victim model’s output behaviors.

\iffalse

\vspace {-1pt} \noindent\textbf{Contributions.} Our main contributions are outlined below: 

\noindent$\bullet$ To the best of our knowledge, we are the first to conduct model stealing attacks against multi-exit networks and the first to explore how to steal output strategies.

%\vspace {3pt}
\noindent$\bullet$  
We introduce performance loss and strategy loss to supervise the extracted model's training. Additionally, we design a novel output strategy search algorithm to maximize the consistency between the victim model and the extracted model's output behavior. Through experiments on multiple mainstream multi-exit networks and benchmark datasets, we throughly demonstrates the effectiveness of our method.

%\noindent$\bullet$
%We conducted experiments on multiple mainstream multi-exit networks and benchmark datasets, thoroughly demonstrating the effectiveness of our method. In addition, we have proposed a feasible defense method and preliminarily verified its validity through experiments.

\vspace{-8pt}

\fi

\vspace{-8pt}
\section{Method}
\label{sec:method}
%-------------------------------------------------------------------------------
%In this section, we propose a novel model stealing attack tailored for multi-exit networks, enabling the theft of both the model's classification function and output strategy. 

\subsection{Threat Model}
\label{subsec:threat-model}
 
In this paper, we assume the attacker has no knowledge of the victim model's internal structure and can only access it in a black-box manner. The attacker can collect a task-irrelevant dataset to query the victim model and fine-tune a pre-trained model to obtain the extracted model. It is worth mentioning that the structure of the extracted model and the victim model do not need to be the same. For example, the victim model’s backbone can be based on VGG, while the extracted model can be built on ResNet.

%\vspace{-6pt}
\subsection{Estimation Stage}
\label{subsec: estimation-stage}
% \vspace{-3pt}

To steal the classification functionality and output strategy of the victim model, we first employ a query dataset to probe the victim model. The output probabilities and exit numbers obtained from the victim model are then used as supervision to train our substitute model. While the output probability of a query sample can be directly obtained, the corresponding exit number cannot be observed from the output results. Therefore, we propose to infer the exit number based on the sample's execution time on the victim model. Since multi-exit networks always output at the first exit that satisfies the output strategy, we can conclude that the runtime of the multi-exit networks is positively correlated with the depth of the exit. Given the significant variations in runtimes across different exits, we propose using the Kernel Density Estimation (KDE) algorithm to analyze the runtime series of all samples. The local minima of the density curve can be interpreted as the temporal boundaries separating different exits.

The KDE algorithm proceeds as follows: First, the entire query dataset is used to query the victim model, obtaining $n$ runtime measurements. These runtimes are then sorted, and the probability density at each runtime point is calculated using the Equation \ref{equation: probability density}. Subsequently, we fit $n$ probability density values to obtain a probability density curve. The $m$ local minima of this curve serve as the runtime thresholds for different exits. By comparing the runtime of each sample to these thresholds, we can determine the corresponding exit number.

\begin{equation}
\label{equation: probability density}
    \hat{f}(x) = \frac{1}{nh}\sum_{i=1}^{n} K\left(\frac{x - x_i}{h}\right)
\end{equation}

\noindent Here, $\hat{f}(x)$ denotes the estimated probability density function at point $x$, $n$ represents the size of query dataset, $K$ is the Gaussian kernel function, and $h$ is the bandwidth.

\iffalse

The principle of Bayesian inference is as follows:
\begin{equation}
\label{equation: Bayesian Inference}
posterior = \frac{prior \times likelihood}{marginal}  
% \vspace{-4pt}
\end{equation} 
Where we use a uniform distribution as the $prior$, indicating that changepoints can empirically occur at any position. Considering that the runtime of each exit approximates a Gaussian distribution, and the intersection between two adjacent distributions has the lowest density, we use the probability density at each time point as the $likelihood$. The $marginal$ is a constant and can be ignored. The resulting $posterior$ represents the probability of each time point being a changepoint. We analyze the runtime vector using BCD and obtain the number and values of changepoints.

\fi

%In our experiments, BCD accurately estimates the number of exits with an accuracy rate of \textcolor{red}{100\%}, and it achieves over \textcolor{red}{99.5\%} accuracy in predicting the exit numbers for the whole test dataset. 

% \vspace{-8pt}
\subsection{Training Stage}
\label{subsec:training-stage}
% \vspace{-4pt}

As mentioned earlier, our attack not only aims to steal the victim model’s classification functionality but also its output strategy. Therefore, our loss function consists of two parts as well. The performance loss guides the extracted model to learn the victim model's classification functionality. The strategy loss helps the extracted model to mimic the output strategy of the victim model, ensuring that query samples are mapped to same exits.

\vspace {2.5pt} 
\noindent\textbf{Performance Loss.}  
The performance loss is used to constrain the classification probabilities of the extracted model to approximate that of the victim model. We achieve this by minimizing the KL-divergence distance across the entire query dataset, as shown in Equation \ref{loss:performance loss}. %\shengzhi{Better to discuss why KL is chosen over other choices.}

% \vspace{-12pt}
\begin{equation}
\label{loss:performance loss}
L_{P}=\min_{\theta_{S}} \mathop{\mathbb{E}}\limits_{x \sim D_{query}} \sum_{i=1}^{K} D_{KL}[S_{i}(x;\theta_{S}),V(x;\theta_{V})]
% \vspace{-4pt}
\end{equation} 
where $i$ represents the exit number, and there are a total of $K$ exits in the multi-exit network. $S_{i}(x;\theta_{S})$ denotes the classification probability from the i-th exit of the extracted model.  $V(x;\theta_{V})$ denotes the classification probability of the victim model's final output. $D_{Query}$ represents the entire dataset used for querying. $\theta_{S}$ and $\theta_{V}$ represent the parameters of the extracted model and the victim model, respectively. $D_{KL}$ refers to the KL-divergence distance.

\vspace {2.5pt} 
\noindent\textbf{Strategy Loss.}  
Given a query sample that is output at the $i$-th exit of the victim model, we aim to ensure that the same sample is also output at the $i$-th exit of the substitute model. Considering that multi-exit networks output at the first exit where confidence exceeds the threshold, we must ensure that the confidence of the given sample is greater than the threshold at the $i$-th exit, while being less than the threshold at all $i-1$ previous exits. This guarantees that the sample will be output by the $i$-th exit of the substitute model and will not be prematurely output at any earlier exit. Our strategy loss is defined by Equation \ref{loss:strategy loss}. 

\vspace{-12pt}
\begin{equation}
\label{loss:strategy loss}
\begin{split}
L_{S}=\min_{\theta_{S}} \sum_{i=1}^{K} \bigg [ \mathop{max}\limits_{x \sim D_{i}} (0,\varphi_{1}-max(S_{i}(x;\theta_{S})))+ \\
\sum_{j=1}^{i-1}\mathop{max}\limits_{x \sim D_{i}} (0,max(S_{j}(x;\theta_{S}))-\varphi_{2}) \bigg ]
\end{split}
\end{equation} 
%\vspace{-12pt}

\noindent $S_{i}(x;\theta_{S})$ denotes the classification probability from the i-th exit of the extracted model.
$max(S_{i}(x;\theta_{S}))$ denotes the maximum value in the classification probability, indicating the model's confidence in the current prediction. $D_{i}$ represents the query samples output from the $i$-th exit of the victim model, and our goal is to train our extracted model to output $D_{i}$ at the $i$-th exit as well. 
%$D_i$ is also referred to as the intended samples for the $i$-th exit of the extracted model. 
%and thus $D_{Query}$ is equal to $D_{1} \cup D_{2} \cup ...\cup D_{K}$.
The first term in the equation aims to maximize the confidence of $D_{i}$ at the $i$-th exit, ensuring that they meet the requirements of the output strategy and are correctly output from the $i$-th exit. The second term in the equation aims to prevent the samples in $D_{i}$ from being prematurely output from any earlier exit. Otherwise, these samples would fail to reach the exits where they should be output. $\varphi_{1}$ and $\varphi_{2}$ are two pre-defined constants\footnote{Through extensive experiments, we find that the values of $\varphi_{1}$ and $\varphi_{2}$ have a minimal impact on the final performance of the model, as long as $\varphi_{1}$ is greater than or equal to $\varphi_{2}$. In our experiments, we set $\varphi_{1}$ to 0.95 and $\varphi_{2}$ to 0.9.}.

\vspace {5pt} 
\subsection{Output Strategy Search.}
\label{subsec: output strategy search}

\iffalse
\begin{figure*}[t]
\centering
\includegraphics[width=1.0\textwidth]{figs/strategy-search.png}
\caption{Output Strategy Search Algorithm.
}
\label{fig:strategy_search}
\vspace{-15pt}
\end{figure*}
%\vspace*{1cm}
%\vspace*{-1cm}
\fi

In practice, simultaneously optimizing performance loss and strategy loss during training can lead to mutual influences. As a result, the final outputs of the substitute model cannot perfectly align with the victim model on all the query samples. Therefore, we aim to search for an optimal set of thresholds as the output strategy, thereby maximizing the output consistency between the extracted model and the victim model.

Specifically, we first use the substitute model to predict all query samples without early output. This means the substitute model outputs multiple predictions for each sample, with each prediction corresponding to one exit. Assuming there are $n$ samples in the query dataset, then each exit has $n$ corresponding confidences. We divide these $n$ confidences for each exit into two sets, $Conf_i$ and $Conf_{\widetilde{i}}$. $Conf_i$ includes the confidence scores of samples that are expected to be output at the $i$-th exit, as they were output from the same exit in the victim model. $Conf_{\widetilde{i}}$ contains the confidences of samples that should not be output from the $i$-th exit. 
The final output strategy $T = \{T_1, T_2, \ldots, T_K\}$ consists of a set of thresholds, one for each of the $K$ exits. The candidate values for $T_i$ are the $n$ confidence values associated with the $i$-th exit. Consequently, there are up to $n^K$ candidate output strategies. In practice, the number of candidate output strategies generated is significantly reduced. This is because $T_i$ exclusively selects confidence scores within the interval $[\min(Conf_i), \max(Conf_{\widetilde{i}})]$ as candidate threshold. Any threshold outside this range would inevitably lead to a higher error rate. 
%Our final output strategy is determined based on Equation \ref{equation: strategy search}.
%For each exit, we iterate over the $n$ confidences and determine the optimal threshold $T_i$ using Equation XX. The optimal threshold maximizes the number of correctly classified samples. By repeating this process for all exit, we construct the final output strategy $T=\{T_1,T_2,\ldots,T_K\}$.

% \vspace{-12pt}
\begin{equation}
\label{equation: strategy search}
T = \underset{ T_{cand}}{argmax} \enspace \sum_{i=1}^{K} \mathbb{I} \{ (Conf_{i} > T_i)  \cup  (Conf_{\widetilde{i}} < T_i) \} 
% \vspace{-4pt}
\end{equation} 

\noindent Here, $T_{cand}$ denote the set of all candidate output strategies. The indicator function $\mathbb{I}$ is used to count the number of elements satisfying subsequent conditions. Equation \ref{equation: strategy search} enables us to search for an output strategy that maximizes the output consistency between the substitute model and the victim model.

\section{Evaluation}
\label{sec:Evaluation}

%HuFu%

\vspace {1pt} \noindent\textbf{Model Structure.} We conduct experiments using the widely adopted multi-exit networks SDN \cite{SDN}. Furthermore, since SDN is implemented by placing multiple classifiers on the backbone network, we consider three different backbone network architectures for it: VGG \cite{VGG}, ResNet \cite{ResNet}, and MobileNet \cite{MobileNet}. In our experiments, we followed the structure and training process set forth in the original paper for the victim model. 
%From Table \ref{table: experiments on different architectures}, it can be observed that the differences in structure between victim models and extracted models have minimal impact on the attack effectiveness. Therefore, For ease of comparison, our substitute model and the target model adopt the same backbone structure. 
Moreover, since the attacker is unaware of the exact locations of the exits on the backbone network, they first estimate the number of exits using KDE and then evenly partition the backbone network of the extracted model to place the classifiers.

\noindent\textbf{Datasets.} 
We employ five mainstream datasets, including CIFAR10 \cite{cifar10}, GTSRB \cite{gtsrb}, SVHN \cite{svhn}, FashionMNIST \cite{fmnist} and TinyImageNet. Among these, the first four are used to train victim models, while the last one is utilized as an unrelated dataset for querying purposes. As stated in the threat model, attackers do not have access to the victim model's training or testing dataset and can only launch attacks using task-irrelevant datasets. 
%For example, to steal the victim model performing the CIFAR10 task, attackers query 100,000 images from TinyImageNet. Among these, 90,000 images are used to train the extracted model, while the remaining 10,000 images are utilized to search for the optimal output strategy.

\noindent\textbf{Baselines.} 
In addition to directly querying with auxiliary datasets, model stealing can also be achieved using random noise and synthetic data. The construction of the query dataset has a significant impact on the effectiveness of model stealing. For a thorough comparison, we select the most representative works from each type of auxiliary dataset as baselines, including KnockOff (Knock)\cite{KnockoffNets}, Noise\cite{Noise}, and ES attack (ESA)\cite{Es-attack}. %Specifically, we first construct auxiliary datasets according to the baseline attacks. Then, we query the victim model to obtain pseudo-labels and train the extracted model following the conventional training methods for multi-exit networks.

\noindent\textbf{Metrics.}We utilized three metrics to evaluate the effectiveness of our method: Accuracy (ACC), Closeness (CLO), and Computation Cost (CC). 

\noindent$\bullet$\textit{ ACC} represents the model's accuracy on the test dataset and is primarily used to measure the model's classification performance. 

\noindent$\bullet$\textit{ CC} quantifies the computational cost of the model across the entire test dataset and is expressed in $10^9$ FLOPs (floating-point operations), serving as an indicator of computational efficiency.

\noindent$\bullet$\textit{ CLO} is a metric used to assess the output consistency between the substitute model and the victim model. Specifically, CLO measures the proportion of samples in the query dataset that produce identical outputs in both models. 

%It is formally defined as follows:
%\begin{equation}
%\label{equation: CLO}
%CLO = \frac{1}{n} \sum_{i=1}^{K} \mathbb{I} \{ (Conf_{i} > T_i)  \cup  (Conf_{\widetilde{i}} < T_i) \} 
% \vspace{-4pt}
%\end{equation} 

\begin{table*}[ht]
\centering
\footnotesize
\caption{Model Performance.}
\label{table: experiments on performance}
\begin{tabular}{m{0.65cm}
<{\centering}|m{0.65cm}
<{\centering}|m{0.7cm}
<{\centering}|m{0.7cm}
<{\centering}|m{1.25cm}
<{\centering}|m{0.7cm}
<{\centering}|m{0.7cm}
<{\centering}|m{1.25cm}
<{\centering}|m{0.7cm}
<{\centering}|m{0.7cm}
<{\centering}|m{1.35cm}
<{\centering}|m{0.7cm}
<{\centering}|m{0.7cm}
<{\centering}|m{1.25cm}
<{\centering}
}

\hline
\multicolumn{2}{c|}{\multirow{2}{*}{\textbf{Models}}} & \multicolumn{3}{c|}{\textbf{CIFAR10}} & \multicolumn{3}{c|}{\textbf{GTSRB}} & \multicolumn{3}{c|}{\textbf{SVHN}} & \multicolumn{3}{c}{\textbf{FMNIST}}\\ 
\cline{3-14}
\multicolumn{2}{c|}{}   &  \textbf{ACC} & \textbf{CLO} & \textbf{CC} & \textbf{ACC} & \textbf{CLO} & \textbf{CC} & \textbf{ACC} & \textbf{CLO} & \textbf{CC} & \textbf{ACC} & \textbf{CLO} & \textbf{CC}   \\ 
\hline

\multirow{5}{*}{VGG} &  Victim & 91.75\% & $-$ & 3133 & 97.13\% & $-$ & 2823 & 95.68\% & $-$ & 8022 & 94.40\% & $-$ & 2771 \\
\cline{2-14}
 & Knock & 88.25\%   & 31.09\%  & 4432(1.41X)  & 95.78\% & 24.40\% & {4868(1.72X)}& 91.99\%   & 23.01\%  & 11787(1.47X)  & 91.48\% & 29.75\% & {3760(1.36X)}\\
 \cline{2-14}
  & Noise    & 33.74\%   & 21.34\%  & 4887(1.56X)  & 45.21\% & 16.78\% & {4008(1.42X)} & 50.03\%   & 38.12\%  & 13156(1.64X)  & 38.56\% & 30.45\% & {3657(1.32X)}\\
 \cline{2-14}
 & ESA    & 82.43\%   & 27.45\%  & 4542(1.45X)  & 86.17\% & 18.76\% & {4742(1.68X)}& 87.29\%   & 32.89\%  & 10990(1.37X)  & 90.92\% & 20.34\% & {3574(1.29X)}\\
 \cline{2-14}
 &  Ours   & 88.33\%   & 63.27\%  & 3258(1.04X)  & 95.42\% & 71.76\% & {3257(1.15X)} & 91.41\%   & 64.85\%  & 7848(0.98X)  & 91.69\% & 57.49\% & {3214(1.16X)}\\
\hline

\multirow{5}{*}{ResNet} &  Victim   & 89.39\%   & $-$  & 1519  & 96.66\% & $-$ & 1141 & 95.11\%   & $-$  & 3633  & 93.34\% & $-$ & 1158\\
\cline{2-14}
 & Knock    & 86.22\%   & 34.69\%  & 1937(1.28X)  & 92.19\% & 18.18\% & {2050(1.80X)} & 92.34\%   & 32.77\%  & 4871(1.34X)  & 90.81\% & 18.87\% & {1650(1.42X)}\\
 \cline{2-14}
  & Noise    & 31.89\%   & 12.89\%  & 2263(1.49X)  & 53.12\% & 25.67\% & {1848(1.62X)} & 42.45\%   & 35.21\%  & 5340(1.47X)  & 34.28\% & 18.56\% & {1598(1.38X)}\\
 \cline{2-14}
  & ESA    & 83.74\%   & 38.12\%  & 2369(1.56X)  & 85.65\% & 15.67\% & {1837(1.61X)} & 89.11\%   & 23.98\%  & 5122(1.41X)  & 88.02\% & 30.21\% & {1655(1.43X)}\\
 \cline{2-14}
 &  Ours   & 86.88\%   & 58.90\%  & 1558(1.03X)  & 92.00\% & 64.30\% & {1374(1.20X)}  & 92.53\%   & 56.39\%  & 3826(1.05X)  & 90.26\% & 51.65\% & {1330(1.15X)}\\
\cline{1-14}

 \multirow{5}{*}{Mobile} &  Victim   & 88.75\%   & $-$  & 3737  & 96.77\% & $-$ & 3203 & 93.72\%   & $-$  & 9178  & 93.08\% & $-$ & 3346\\
\cline{2-14}
 & Knock    & 83.35\%   & 41.92\%  & 5199(1.39X)  & 91.35\% & 30.77\% & {4788(1.59X)} & 90.52\%   & 41.86\%  & 12952(1.41X)  & 89.43\% & 27.19\% & {4339(1.29X)}\\
 \cline{2-14}
  & Noise    & 49.76\%   & 31.09\%  & 5717(1.53X)  & 52.01\% & 14.23\% & {4676(1.46X)} & 36.45\%   & 29.76\%  & 12390(1.35X)  & 30.87\% & 33.89\% & {4918(1.47X)}\\
 \cline{2-14}
  & ESA    & 82.79\%   & 17.54\%  & 5381(1.44X)  & 80.34\% & 25.87\% & {4932(1.54X)} & 86.44\%   & 19.23\%  & 12482(1.36X)  & 84.58\% & 28.65\% & {5019(1.50X)}\\
 \cline{2-14}
 &  Ours   & 83.39\%   & 59.72\%  & 3914(1.05X) & 91.87\% & 69.42\% & {4095(1.28X)} & 89.37\%   & 62.44\%  & 10212(1.11X) & 89.37\% & 59.60\% & {3580(1.07X)} \\
\cline{1-14}
\hline
\end{tabular}
\end{table*}

\begin{table}[ht]
\footnotesize
\caption{The Prediction Accuracy of Exit Numbers}
\vspace*{-0.5cm}
\label{table: KDE accuracy}
\begin{center}
\begin{tabular}{m{1.0cm}
<{\centering}m{1.4cm}
<{\centering}m{1.4cm}
<{\centering}m{1.4cm}
<{\centering}S[table-format=2.3, table-column-width=1.3cm]}
\hline
\multicolumn{1}{c}{\textbf{Extracted}}&
\multicolumn{1}{c}{\textbf{CIFAR10}}&
\multicolumn{1}{c}{\textbf{GTSRB}} & 
\multicolumn{1}{c}{\textbf{SVHN}}&
\multicolumn{1}{c}{\textbf{FMNIST}} \\
\hline\hline

{VGG} & {0.998$\pm$1e-6} & {0.999$\pm$2e-5} & {0.999$\pm$1e-4}&{0.998$\pm$3e-6} \\
\hline

{ResNet} & {1.0$\pm$0.0} & {0.998$\pm$3e-5} & {1.0$\pm$0.0}&{0.999$\pm$1e-6} \\
\hline

{Mobile} & {0.998$\pm$2e-5} & {1.0$\pm$0.0} & {0.999$\pm$6e-7}&{0.998$\pm$1e-5} \\
\hline

\end{tabular}
\end{center}
\vspace{-15pt}
\end{table}

\subsection{Performance Evaluation}

As shown in Table \ref{table: experiments on performance}, we first evaluated the effectiveness of baselines and our approach on four datasets. Through testing on the victim model's test dataset, we found that the extracted model trained by our method always achieves the closest accuracy to the victim model. This implies that our method can effectively learn the classification functionality of the victim model. In terms of computational efficiency, our method incurs a comparable computational cost to the victim model, significantly lower than all baseline methods. For instance, with the VGG on CIFAR10, our method incurs only a 4\% additional computational overhead compared to the victim model, while the three baseline methods introduce computational overheads that are 10 times, 14 times, and 11 times greater than ours. Moreover, our method's CLO also significantly exceeds all baseline methods, indicating a stronger consistency in output behavior between our extracted model and the victim model. These experimental results fully demonstrate that our method can successfully learn the output strategies of the victim model. 
%Due to space constraints, we only display the results for CIFAR10 and GTSRB in Table \ref{table: experiments on performance}. The results for SVHN and FashionMNIST, which are similar to those of CIFAR10 and GTSRB, have been attached in the appendix.

\subsection{Accuracy on Exit Prediction}

As shown in Table \ref{table: KDE accuracy}, our approach achieves nearly 100\% accuracy in predicting query sample's exits across various datasets and network architectures. This indicates that the runtime of a sample is indeed positively correlated with the depth of the corresponding exit, and KDE can effectively capture the temporal differences between different exits.

\subsection{Sensitivity Analysis of Architectures and parameters}

%\vspace {5pt} 
\noindent\textbf{Numbers of Exits.} The difficulty of model stealing may vary depending on the number of exits in the victim model. Here, we conducted experiments on four datasets, CIFAR10, GTSRB, SVHN, and FashionMNIST, and observed the attack effectiveness of our method when the victim models have 3, 4, 5, and 6 exits. Our experiments reveal that the number of exits has a very small impact on the attack effectiveness of our method, and the extracted models with different numbers of exits always exhibit very similar accuracy and closeness. 
Additionally, as the number of exits increases, we observe a slight decrease in the total computational cost. This is mainly because more samples output at the shallow exits, with only a few complex samples reaching the deep exits.

\begin{table}[ht]
\footnotesize
\caption{Extracted Models with Different Backbone Architectures}
\vspace*{-0.5cm}
\label{table: experiments on different architectures}
\begin{center}
\begin{tabular}{m{1.5cm}
<{\centering}m{1.2cm}
<{\centering}m{1.1cm}
<{\centering}m{1.1cm}
<{\centering}S[table-format=2.3, table-column-width=1.5cm]}
\hline
\multicolumn{1}{c}{\textbf{Extracted}}&
\multicolumn{1}{c}{\textbf{Victim}}&
\multicolumn{1}{c}{\textbf{ACC}} & 
\multicolumn{1}{c}{\textbf{CLO}}&
\multicolumn{1}{c}{\textbf{CC}} \\
\hline\hline
\multirow{3}{*}{VGG} 
& {VGG} & {88.33\%} & {63.27\%}&{3258(1.04X)} \\
& {ResNet} & {86.47\%} & {58.86\%}&{3383(1.08X)} \\
&{MobileNet}&{86.54\%} & {61.70\%} & {3289(1.05X)}\\
\hline

\multirow{3}{*}{ResNet} & {VGG} & {87.70\%} & {53.59\%}&{1610(1.06X)} \\
& {ResNet} & {88.68\%} & {64.24\%}&{1558(1.03X)} \\
&{MobileNet}&{86.76\%} & {55.52\%} & {1595(1.05X)}\\

\hline
\multirow{2}{*}{Mobile} & {VGG} & {82.68\%} & {50.04\%}&{3961(1.06X)} \\
&{ResNet}&{82.49\%} & {57.81\%} & {4036(1.08X)}\\
& {MobileNet} & {83.39\%} & {59.72\%}&{3923(1.05X)} \\
\hline
\end{tabular}
\end{center}
\vspace{-15pt}
\end{table}

%\vspace {5pt} 
\noindent\textbf{Backbone Architectures.} 
Because attackers are unaware of the victim model's network architecture in black-box scenarios, it is necessary to evaluate the impact of using different structures in extracted models on attack effectiveness. We conducted experiments using a 4-exit model structure on the CIFAR10 dataset. As shown in Table \ref{table: experiments on different architectures}, When using the same backbone structure as the victim model, the extracted model often achieves the highest accuracy and closeness. When using a different structure, the accuracy and computational cost of the extracted model averagely decrease by  1.36\% and 2.16\% respectively, still achieving good performance. This indicates that the choice of the extracted model’s structure has little impact on our method.

\vspace{-10pt}

\section{Conclusion}
\label{sec:conclusion}

In this paper, we propose a novel model stealing attack targeting multi-exit networks, which extracts both the model functionality and output strategy.  We employ KDE to analyze the target model's output strategy and use performance loss and strategy loss to guide the training of the extracted model. 
Furthermore, we designed a novel output strategy search algorithm that can find the optimal output strategy to maximize the consistency between the victim model and the extracted model's outputs.
%Through experiments on multiple mainstream multi-exit networks and benchmark datasets, our method always achieves similar classification performance and computational cost with the victim models, 
% Compared to extracted models trained by traditional methods, our extracted models have a much higher computational efficiency and closeness, 
%which fully demonstrates the effectiveness of our method.
Various experimental results thoroughly demonstrate the effectiveness of our method. 

%\shengzhi{What are those numbers in red after each reference?} ==>
%\lipan{This is a specific citation format used in CVPR, indicating on which page this work is cited.}

% \vspace{-20pt}
% \section*{Acknowledgment}

% We thank the reviewers for their constructive feedback. The IIE authors are supported in part by Beijing Natural Science Foundation (No.M22004), NSFC (92270204, 62302497), Youth Innovation Promotion Association CAS and a research grant from Huawei.

%\bibliographystyle{IEEEbib}
%\bibliography{strings,refs}

{\small
\bibliographystyle{IEEEbib}
\bibliography{references}
}

\end{document}